\documentclass[sigconf]{acmart}

\AtBeginDocument{%
  \providecommand\BibTeX{{%
    \normalfont B\kern-0.5em{\scshape i\kern-0.25em b}\kern-0.8em\TeX}}}

\usepackage[colorinlistoftodos]{todonotes}
\usepackage{algorithmic}
\usepackage{graphicx}
\usepackage{textcomp}
\usepackage{xcolor}
\usepackage{booktabs}
\usepackage{balance}
\usepackage{hyperref}
\usepackage{soul}
\usepackage[normalem]{ulem}
\usepackage{framed}
\usepackage{orcidlink}

\usepackage{enumitem}
\setlist[itemize]{noitemsep, topsep=0pt}

\setcopyright{none}
\renewcommand\footnotetextcopyrightpermission[1]{}
\begin{document}

\title{\textcolor{black}{Investigating the Efficacy of Large Language Models for Code Clone Detection}}
%
%
%
%

\author{Mohamad Khajezade}
\affiliation{%
  \institution{University of British Columbia}
  \streetaddress{Address Line}
  \city{Kelowna}
  \state{B.C.}
  \postcode{Zip Code}
  \country{Canada}
}
\email{khajezad@mail.ubc.ca}

\author{Jie JW Wu\orcidlink{0000-0002-7895-2023}}
\affiliation{%
  \institution{University of British Columbia}
  \streetaddress{Address Line}
  \city{Kelowna}
  \state{B.C.}
  \postcode{Zip Code}
  \country{Canada}
}
\email{jie.jw.wu@ubc.ca}

\author{Fatemeh Hendijani Fard}
\affiliation{%
  \institution{University of British Columbia}
  \streetaddress{Address Line}
  \city{Kelowna}
  \state{B.C.}
  \postcode{Zip Code}
  \country{Canada}
}
\email{fatemeh.fard@ubc.ca}

\author{Gema Rodríguez-Pérez}
\affiliation{%
  \institution{University of British Columbia}
  \streetaddress{Address Line}
  \city{Kelowna}
  \state{B.C.}
  \postcode{Zip Code}
  \country{Canada}
}
\email{gema.rodriguezperez@ubc.ca}

\author{Mohamed Sami Shehata}
\affiliation{%
  \institution{University of British Columbia}
  \streetaddress{Address Line}
  \city{Kelowna}
  \state{B.C.}
  \postcode{Zip Code}
  \country{Canada}
}
\email{mohamed.sami.shehata@ubc.ca}
\begin{abstract}

Large Language Models (LLMs) have demonstrated remarkable success in various natural language processing and software engineering tasks, such as code generation. The LLMs are mainly utilized in the prompt-based zero/few-shot paradigm to guide the model in accomplishing the task.  
GPT-based models are one of the popular ones studied for tasks such as code comment generation or test generation. These tasks are `generative' tasks. However, there is limited research on the usage of LLMs for `non-generative' tasks such as classification using the prompt-based paradigm.  
In this preliminary exploratory study, we investigated the applicability of LLMs for Code Clone Detection (CCD), a non-generative task. 
By building a mono-lingual and cross-lingual CCD dataset derived from CodeNet, we first investigated two different prompts using ChatGPT to detect \textcolor{black}{Type-4} code clones in Java-Java and Java-Ruby pairs in a zero-shot setting. We \textcolor{black}{then} conducted an analysis to understand the strengths and weaknesses of ChatGPT in CCD. 
ChatGPT surpasses the baselines in cross-language CCD \textcolor{black}{attaining an F1-score of 0.877 } and achieves comparable performance to fully fine-tuned models for mono-lingual CCD, \textcolor{black}{with an F1-score of 0.878}. Also, the \textcolor{black}{prompt and the} difficulty level of the problems has an impact on the performance of ChatGPT. \textcolor{black}{Finally,} we provide insights and future directions based on our initial analysis\footnote{Our code and data is open-sourced at \url{https://github.com/mkhfring/largeLanguageModels}}. 

\end{abstract}


\maketitle
\section{Introduction}\label{intro}
Code clones, the functionally identical code fragments, can cause buggy code to propagate through the whole project~\cite{ain2019systematic,lei2022deep}. Code Clone Detection (CCD) is necessary to develop and maintain source code, and is beneficial for various applications, including finding library candidates, code comprehension, finding harmful software, and plagiarism detection. Code clone is categorized into four types depending on the similarity measures: 1) \textbf{Type-1 clones}:  \textcolor{black}{Code fragments that are identical in terms of their code and structure, except for variations in whitespace, comments, and layout}. 2) \textbf{Type-2 clones:} \textcolor{black}{ Code fragments that are syntactically identical except for variations in identifiers, literals, types, whitespace, layout, and comments.} 3) \textbf{Type-3 clones}: Code fragments that are mostly similar, but can have differences in terms of statements, in addition to the variations in identifiers, literals, types, whitespace, layout, and comments. 4) \textbf{Type-4 clones}: \textcolor{black}{Code snippets that are completely different syntactically and structurally but share the same functionality}. Among them, Type-4 clones are the most challenging to detect given their different syntactic variants in implementation and thus require a high degree of reasoning ability~\cite{ain2019systematic}.

\textcolor{black}{Different types of approaches~\cite{ain2019systematic} have been proposed for CCD, including text-based approaches, lexical approaches, syntactic approaches, and semantic approaches.} In recent years, deep learning approaches have been a major approach for different software engineering tasks, such as clone detection. \textcolor{black}{The deep learning-based code clone detections leverage deep models to learn the inherent similarities of clones~\cite{lei2022deep}}. These code clone detectors can effectively detect different clone types, such as Type-3 and Type-4. However, \textcolor{black}{these methods have been proven to perform poorly with unseen problems and out-of-distribution data \cite{9678907,sonnekalb2022generalizability}. Additionally, these models are not effective for cross-language code clone detection, where code snippets are implemented in different languages \cite{tao2022c4}.}

Recently, Large Language Models (LLM) have been a prominent approach due to their emergent abilities \cite{brown2020language,min2021recent,weng2022identification,xu2022systematic,brown2020language}. LLMs are models trained on a large number of parameters (e.g., GPT-3 has 175 billion parameters \cite{brown2020language}) in an unsupervised manner and have had a profound impact on multiple fields, including natural language processing \cite{min2021recent}, machine learning \cite{weng2022identification}, and software development \cite{xu2022systematic}. These models can analyze large volumes of textual data, allowing for accurate predictions and the generation of different contents in a zero-shot manner, i.e., without the need to employ any labeled data \cite{brown2020language}. However, it remains unclear what values can LLMs provide to the problem of CCD, especially Type-4. 
To address this problem, our goal in this paper is to investigate and understand the performance and limitations of using LLM for CCD. We chose Type-4 clone because \textcolor{black}{detecting this type of code clone is challenging. Previous models have found it difficult to address Type-4 code clones, especially in the contexts of few-shot learning and cross-language clone detection \cite{9678907}.  
} \textcolor{black}{We believe the reasoning capabilities of large language models provide a promising approach for Type-4 clones}. 

\textcolor{black}{In this paper, we first investigated two different prompts using ChatGPT to detect code clones in Java-Java and Java-Ruby pairs in zero-shot setting by building a mono-lingual and cross-lingual CCD dataset derived from CodeNet.} Furthermore, we conducted an analysis to understand the strengths and weaknesses of ChatGPT in CCD. Our main finding is that \textcolor{black}{ChatGPT surpasses the baselines in cross-language CCD and achieves comparable performance to fully fine-tuned models for mono-lingual CCD. } \textcolor{black}{
The F1 performance of ChatGPT for cross-language CCD is 0.877, in contrast to baselines such as CodeBERT, which achieves an F1-score of 0.663. 
} 
{Additionally, the prompt and the difficulty level of the problems have an impact on the performance of ChatGPT.}

\textcolor{black}{
\textbf{Difference between this work and existing studies:} 
Though there exist multiple studies on using LLMs for various code-related and software engineering tasks~\cite{trummer2022codexdb,nashid2023retrieval,bareiss2022code}, the number of works on using LLMs for code clones is limited. 
The only related works are~\cite{roy2023unveiling, dou2023towards}. 
The former explores the application of GPT-3 in generating semantic clones across languages, which is not detecting clones. 
The latter assesses the performance of different LLMs in CCD for mono-lingual setting. Our work is different from these works, as we focus on mono- and cross-language CCD. Additionally, we are the first to investigate the complexity and difficulty level of the problems for CCD using LLMs. }

\section{Investigations}
We investigated the efficacy of LLM for CCD by two RQs: 

\textbf{RQ1:} \textit{What is the effect of different prompts to encourage ChatGPT to identify Code Clones?}
Here, we investigate the effect of our designed prompt on the performance of ChatGPT for \textcolor{black}{Type-4} CCD, both uni-language and cross-language. The goal is to investigate to what extent the new prompts can improve performance.

\textbf{RQ2:} \textit{What is the performance of ChatGPT for code clone detection compared to the baselines?}
In this RQ, we are interested in comparing the results of ChatGPT with the baselines to detect code clones. Note that for this experiment, ChatGPT is tested in a zero-shot setting, while the baselines are fully fine-tuned. 

\subsection{Investigations Methodology}

To answer RQ1 and RQ2, we assess the performance of ChatGPT (GPT-3.5-turbo) in zero-shot binary code clone detection with different prompts based on F1-Score, Cyclomatic Complexity and Acceptance Rate. The prompt $P_c$, which  directs ChatGPT to determine whether two code snippets are code clones, is defined as follows:

\begin{equation}
P_c = \{C_{i,j}, C_{k,l}, NL\}
\end{equation}

Here, $i$ and $k$ represent the problem number in the dataset (CodeNet), while $j$ and $l$ denote the submission number within problems $i$ and $k$. If $k=i$, the two code snippets in the prompt are code clones; otherwise, they are non-clone pairs.

\begin{figure}
\centering
\includegraphics[width=0.5\textwidth]{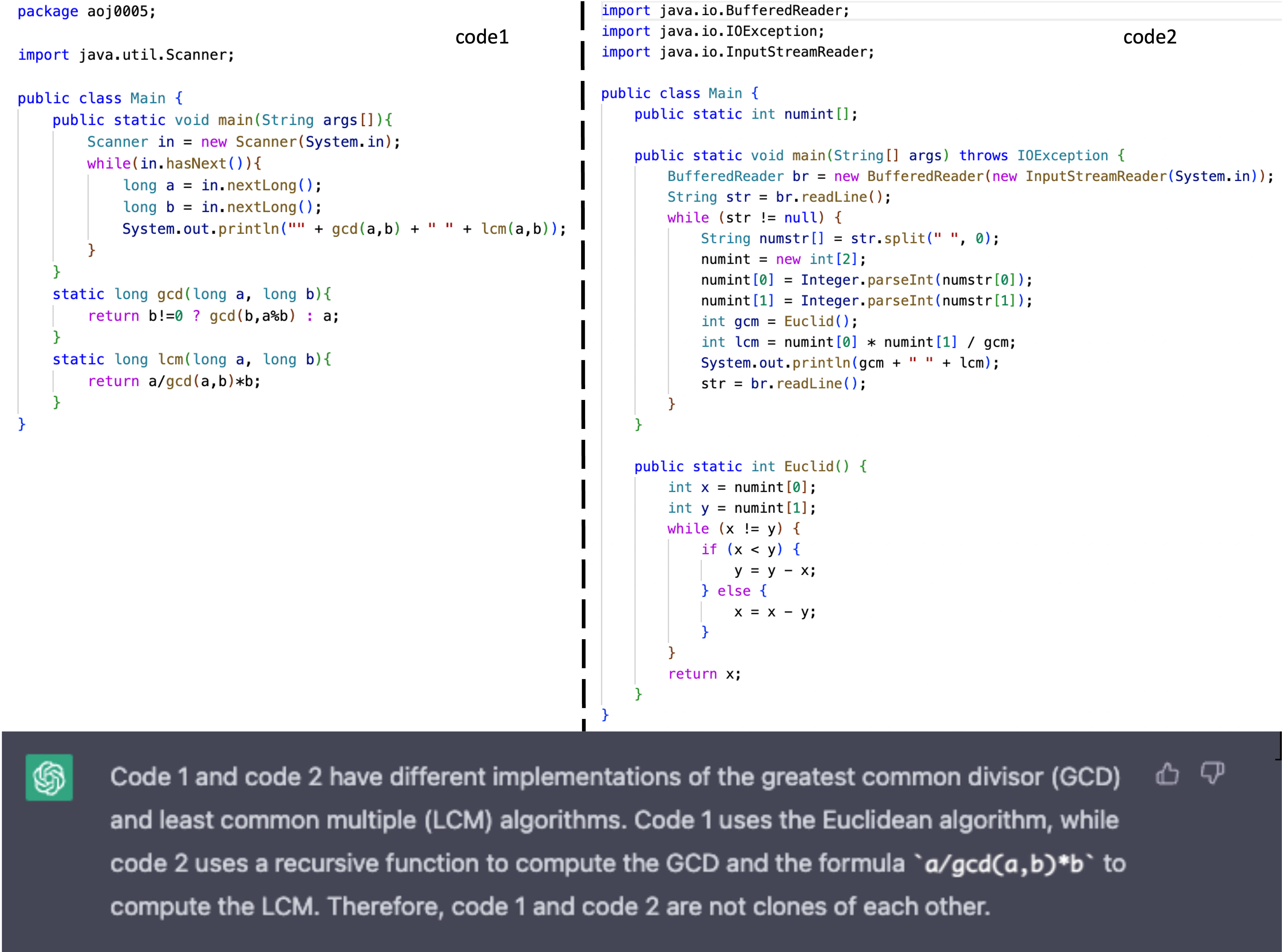}
\caption{Code1 and code2 are two accepted submissions for problem \#5 of CodeNET dataset \cite{puri2021project}. The figure displays the result of querying ChatGPT using a basic prompt asking whether code1 and code2 are code clones. Though ChatGPT correctly identifies the functionality of both scripts, it fails to identify whether these two are clones.}
\label{fig:motivation}
\end{figure}

\begin{figure}[!h]
\centering
\includegraphics[width=0.5\textwidth]{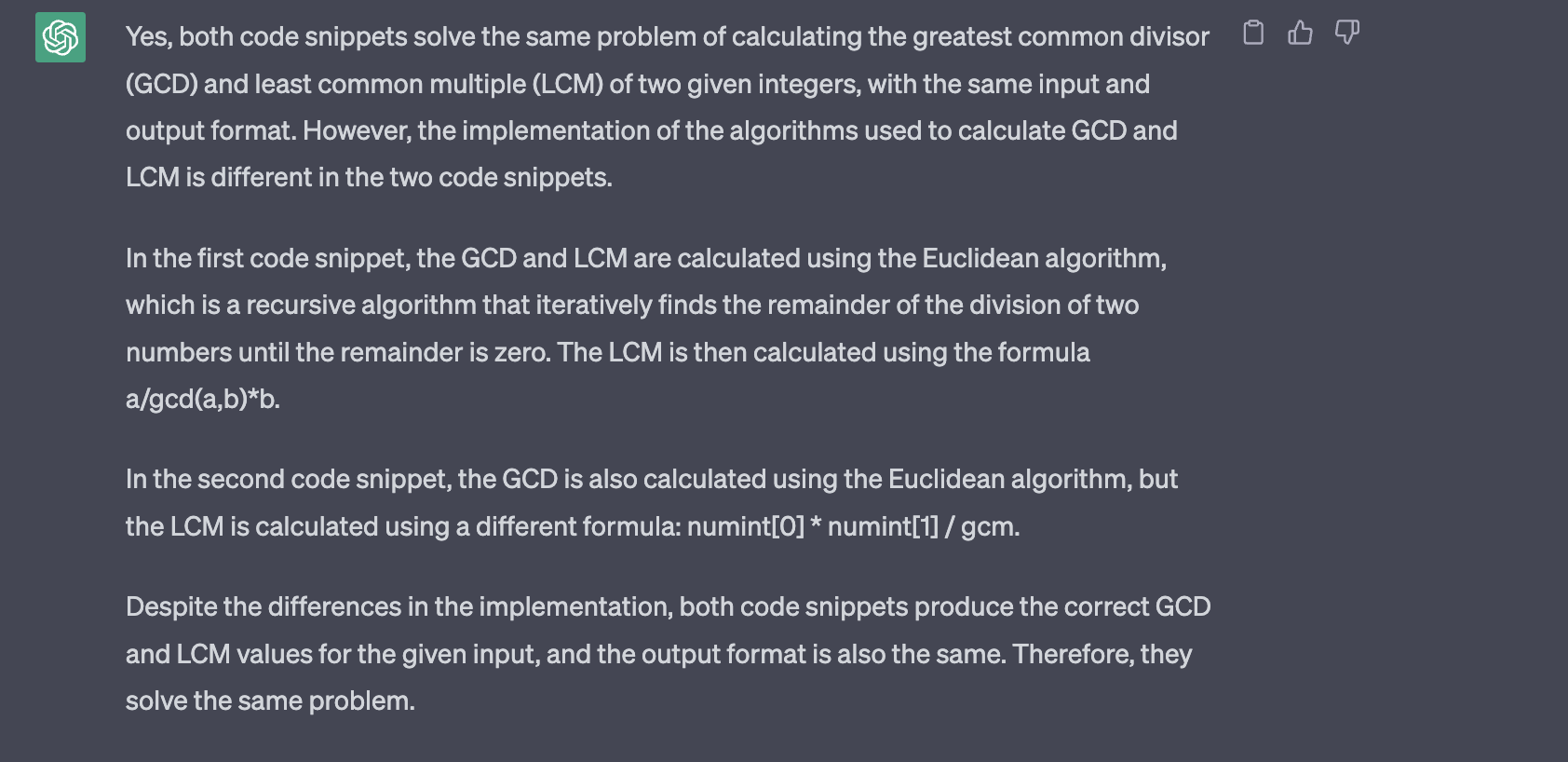}
\caption{Code1 and code2 are two accepted submissions for problem \#5 of CodeNET dataset \cite{puri2021project} as shown in Figure \ref{fig:motivation}. The figure displays the result of querying ChatGPT using our designed prompt when it correctly mentions that the two codes are solving the same problem.}
\label{fig:motivation2}
\end{figure}

\textbf{Prompt Design:} In our investigations of prompt, we found that directly asking ChatGPT whether two codes are clones of each other is ineffective. For example, Figure \ref{fig:motivation} displays two code implementations (code1 and code2) of the Greatest Common Divisor (GCD) and ChatGPT's ability to detect whether the two codes are clones. Despite identifying the intention behind both codes, ChatGPT fails to accurately recognize whether they are clones or not, indicating a misconception in its understanding of code clones. Hence, asking the model whether two given codes are clones does not result in a correct answer. In contrast, Figure \ref{fig:motivation2} presents ChatGPT's results for the same code snippets when using the prompts developed in this paper. As depicted, ChatGPT accurately classifies these code segments as code clones. 


Thus, our prompt is designed to ask ChatGPT to evaluate whether $C_{i, j}$ and $C_{k, l}$ intend to solve the same problem. In other words, ChatGPT should determine if $i=k$. Subsequently, if $i=k$, ChatGPT is tasked with ascertaining whether $C_{i, j}$ and $C_{k, l}$ possess the same inputs and outputs. In our experiments, we limited ChatGPT to output either "Yes" for clone pairs or "No" for non-clone pairs. Then, the results mapped to 1 for code clones and 0 otherwise.  \textcolor{black}{
The final prompt will be as follows:}
\begin{framed}
\textcolor{gray}{
\\
"""\\
\{\textbf{code1}\},
\{\textbf{code2}\}, \\
Do code 1 and code 2 solve identical problems with the same inputs and outputs? answer with yes or no and no explanation.\\
"""}
\end{framed}
\textcolor{black}{Note that code1 and code2 are the requested codes in the above prompt.
}

\subsection{Evaluation Metrics}
We adopted the following three metrics in the evaluation:

\textcolor{black}{
\textbf{F1-Score}
The F1-Score, also referred to as the F-Score or F-Measure, is adopted to evaluate the performance of models for binary CCD, since it's a widely used performance metric for assessing the accuracy of a binary classifier \cite{derczynski2016complementarity}. 
\\
\textbf{Cyclomatic Complexity (CC)}
We employed the average Cyclomatic Complexity \cite{ebert2016cyclomatic} to determine the difficulty of each problem. Cyclomatic Complexity \cite{ebert2016cyclomatic} is a software metric widely used to measure the complexity of a given piece of code, based on the control flow graph of the code. A higher Cyclomatic Complexity indicates more complex code, often making it more challenging to maintain, understand, and test \cite{moreno2016comparing}. \\
\textbf{Acceptance Rate}
We computed this metric for each selected problem in the CodeNet dataset to analyze the difficulty level of each problem, based on its metadata. The metric is calculated using the equation $\text{acceptance\_rate}_{p_i} = \frac{C{p_i}}{A{p_i}}$, where $p_i$ represents a selected problem from the CodeNet dataset, $A_{i}$ denotes the total number of submissions for $p_i$, and $C_{i}$ signifies the total number of correct submissions for $p_i$. The reason we use this metric is because it has been used in previous studies to assess the difficulty of programming problems \cite{intisar2018cluster} and to enable advanced deep-learning systems to categorize problems based on their level of difficulty \cite{intisar2018cluster}. 
}
\subsection{Clone Dataset (Type-4)}
\textcolor{black}{For our code clone detection study, we used the Project CodeNet dataset \cite{puri2021project}, containing 14 million code samples from 4,000 coding problems in over 50 languages, primarily C++, C, Python, and Java. These samples include accepted and rejected solutions, along with metadata like code size and acceptance status. Tools on CodeNet's GitHub repository~\footnote{https://github.com/IBM/Project\_CodeNet} enable the transformation of these samples into token sequences or parse trees.
We \textcolor{black}{sampled} two datasets: one for Java-Java (CCD\textsubscript{JJ}) and anothder for Java-Ruby (XCCD\textsubscript{JR}) pairs, to explore uni-language and cross-language code clone detection. Java was chosen for its widespread use and Ruby for its syntactical difference and lower resource availability. Each dataset comprises 1,000 samples (500 positive, 500 negative pairs) randomly selected from 100 problems. Positive pairs were from the same problem, while negative pairs were from different problems, ensuring a balanced and representative selection.
We used CodeNet's metadata to filter accepted solutions. Our selection methodology was designed to provide a valid comparison of results. We also chose CodeNet over other datasets like CodeSearchNet, hypothesizing that ChatGPT might be less familiar with CodeNet, which requires preprocessing, potentially leading to more insightful results about the model's capabilities. 
It is worth noting that the research did not use the standard datasets for code clone detection for two reasons: First, for cross-language code clone detection, we need datasets comprising the same problems implemented in different languages. Second, we require metadata to compute the difficulty of each problem.}

\textcolor{black}{
\subsection{Baselines}\label{baseline}
For code clone detection, we compare ChatGPT, which uses a zero-shot approach, with 
baselines fully fine-tuned on the CodeNET dataset. These models, widely used in literature \cite{yu2023graph,tao2022c4}, include:\\
\textbf{CodeBERT:} \cite{feng-etal-2020-codebert} A bimodal pre-trained model for program comprehension and natural language tasks, using transformer architecture and a hybrid objective function. It excels in tasks like code search and code comment generation \cite{feng-etal-2020-codebert}, making it a fitting baseline.\\
\textbf{RoBERTa} \cite{liu2019roberta} This enhanced BERT variant focuses on Masked Language Model training, with a larger learning rate, dynamic masking, and training on a broad text corpus. It's chosen for its proven performance in code clone detection \cite{wang2021syncobert,lu2021codexglue} and its presence on the CodeXGLUE leaderboard\footnote{https://microsoft.github.io/CodeXGLUE/}.\\
\textbf{GraphCodeBert} \cite{guo2021graphcodebert} A transformer model that emphasizes data flow over syntactic structures, such as AST, in pre-training. It employs structure-aware approaches for enhanced code representation and is effective in tasks like code clone detection, translation, refinement, and search.\\
\textcolor{black}{It should be noted that we excluded C4 \textcolor{black}{\cite{tao2022c4}} from our baseline for cross-language detection, due to data format constraints.}}
\section{Investigation Results}
\textcolor{black}{Table \ref{tab:different-temps} presents the results of varying 
temperatures in the API call for Java-Java pairs. \textcolor{black}{These results are based on the proposed prompt design as described earlier.} As shown, the F1-Scores for $T=0.1$, $T=0.3$, and $T=0.5$ are $0.852$, $0.855$, and $0.878$, respectively. The optimal performance is achieved at $T=0.3$. Consequently, this temperature is used for Java-Ruby pairs. As displayed in table \ref{tab:different-temps}, the F1-score for Java-Ruby is 0.877. 
}

\begin{table}[]
    \centering
    \caption{The results of using ChatGPT for Java-Java pairs and Java-Ruby pairs from CodeNet with different temperatures }
    \resizebox{0.9\columnwidth}{!}{%
    \begin{tabular}{@{}lccc@{}}\toprule
   Experiment & Precision & Recall & F1-score\\
    \midrule

        Java Codes with T=0.3& 0.784 & 0.997 & 0.878
        \\
        
        Java and Ruby Codes with T=0.3 & 0.796 & 0.977 & 0.877
        \\
        
        Java Codes with T=0.1& 0.887 &  0.855 & 0.852
        \\
        
        Java Codes with T=0.5& 0.889 & 0.858 & 0.855

        \\

    \bottomrule
    \end{tabular}}
    
    \label{tab:different-temps}
\end{table}

\begin{table}[]
    \centering
    \caption{The results for different prompts. Prompt-1 simply asked if two code snippets are code clones. Prompt-2 is our proposed prompt}
    \resizebox{0.8\columnwidth}{!}{%
    \begin{tabular}{@{}lccc@{}}\toprule
   Experiment & Precision & Recall & F1-score\\
    \midrule

        Java-Java Prompt-1& 1.0 & 0.348& 0.517
        \\
        
        Java-Ruby Prompt-1 & 0.948 & 0.745 & 0.834       
        \\
        
        Java-Java Prompt-2 & 0.784 &  0.997 & 0.878
        \\
        
        Java-Ruby Prompt-2 & 0.796 & 0.977 & 0.877

        \\

    \bottomrule
    \end{tabular}}
    
    \label{tab:prompt-compare}
\end{table}

\begin{table}[]
    \centering
    \caption{The performance of the models for Java-Java pairs}
    \resizebox{0.7\columnwidth}{!}{%
    \begin{tabular}{@{}lccc@{}}\toprule
   Baseline & Precision & Recall & F1-score\\
    \midrule

        CodeBERT & 0.912 & 0.881 & 0.896
        \\
        
        RoBERTa & 0.899 & 0.852 & 0.874
        \\
        GraphCodeBERT & \textbf{0.947} & 0.883& \textbf{0.914}
        \\
        \textbf{ChatGPT} & 0.784 & \textbf{0.997} & 0.878

        \\

    \bottomrule
    \end{tabular}}
    
    \label{tab:baseline-compare-java}
\end{table}

\begin{table}[]
    \centering
    \caption{The performance of the models for Java-Ruby pairs }
    \resizebox{0.7\columnwidth}{!}{%
    \begin{tabular}{@{}lccc@{}}\toprule
   Baseline & Precision & Recall & F1-score\\
    \midrule
        CodeBERT & 0.498& 0.988 & 0.663
        \\
        
        RoBERTa &0.486 & \textbf{0.992}  & 0.652
        \\
        GraphCodeBERT &0.50 & 0.991& 0.665
        \\
        
        \textbf{ChatGPT} & \textbf{0.796} & 0.977 & \textbf{0.877}

        \\

    \bottomrule
    \end{tabular}}
    
    \label{tab:baseline-compare-ruby}
\end{table}

\begin{table}[]
    \centering
    \caption{The Acceptance Rate and Cyclomatic Complexity metrics are averaged for misclassified samples shared between Java-Java and Java-Ruby datasets. The positive samples represent the average of clone pairs, while the negative samples indicate the average for non-clone pairs}
    \resizebox{0.9\columnwidth}{!}{%
    \begin{tabular}{@{}lcc@{}}\toprule
    Experiment & Acceptance Rate & Cyclomatic Complexity\\
    \midrule

        Positive Samples & 0.038 & 2.54 
        \\
        
        Negative Samples & 0.035 & 3.16  
        \\
        Selected Problems & 0.058 & 2.98
        \\

    \bottomrule
    \end{tabular}}
    
    \label{tab:complexity}
\end{table}

\begin{figure}[t!]
\centering
\includegraphics[width=0.5\textwidth]{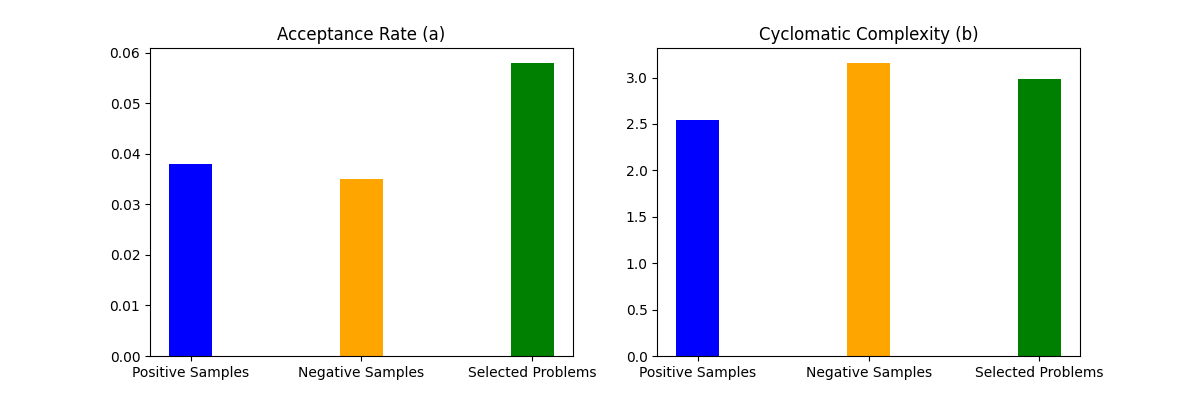}
\caption{a) Comparison of the average Acceptance Rates for positive (\textcolor{blue}{blue}) and negative (\textcolor{orange}{orange}) misclassified samples, and the selected CodeNet problems 
(\textcolor{green}{green}). The lower Acceptance Rate indicates more sophisticated problems. b) Comparison of the average Cyclomatic Complexity for positive (\textcolor{blue}{blue}) misclassified samples and negative (\textcolor{orange}{orange}) mislabeled samples, and the selected programs from CodeNET (\textcolor{green}{green}). A higher CC represents more difficult problems.}
\label{fig:complexity}
\end{figure}

\textcolor{black}{\textbf{RQ1: Effect of different prompts. }
Table \ref{tab:prompt-compare} compares two prompts, \textit{Prompt-1}, which simply asks ChatGPT if two codes are code clones, and \textit{Prompt-2}, which is the prompt proposed in this research. As illustrated, the lowest performance is observed with Prompt-1 for Java-Java pairs, yielding an F1-score of $0.517$, a precision of $1.0$, and a recall of $0.348$. Interestingly, Prompt-1 achieves an F1-score of $0.834$ for Java-Ruby pairs, with a precision of $0.948$ and a recall of $0.745$. The prompt designed in this research, Prompt-2, outperforms Prompt-1 for both Java-Java pairs, with an F1-score of $0.878$, and Java-Ruby pairs, with an F1-score of $0.877$.\\
\textbf{RQ2: Comparing the performances.}
As Prompt-2 achieved the best scores in our first experiment, we compare the results of Prompt-2 with the baselines in this RQ2.
For Java-Ruby pairs, as displayed in Table \ref{tab:baseline-compare-ruby}, ChatGPT delivers the best performance with an F1-score of $0.877$. Consequently, the performance of CodeBERT, RoBERTa, and GraphCodeBERT are $0.663$, $0.652$, and $0.665$, respectively. 
In summary, ChatGPT surpasses other baselines in performance for the Java-Ruby cross-language dataset, achieving an F1-score of 0.877. Conversely, the other baselines yield lower F1-scores in the range of 0.66 for Java-Ruby pairs, with precision and recall metrics indicating that these baselines predominantly generate $0$. However, baselines exhibit improved F1 scores for Java-Java pairs compared to ChatGPT, with scores ranging from 0.896 to 0.914, as opposed to 0.878 for ChatGPT. Notably, ChatGPT accomplishes this performance in a zero-shot manner, while the baselines are fully fine-tuned for the downstream task.}



\section{Discussions}
The results presented in Table \ref{tab:different-temps} suggest that the optimal performance is achieved with a temperature of 0.3. However, as shown, altering the temperature does not significantly impact performance. Another noteworthy observation from Table \ref{tab:different-temps} is that the performance of Java-Java pairs is nearly identical to that of Java-Ruby pairs, even though cross-language clone detection poses a challenge for other baselines. One plausible explanation for this outcome is that ChatGPT struggles with specific problems for both Java-Java and Java-Ruby pairs. 
\textcolor{black}{Considering Tables \ref{tab:baseline-compare-java} and \ref{tab:baseline-compare-ruby}, we observe a performance drop of the baselines for the cross-language setting. For instance, the F1-score of GraphCodeBERT decreases from 0.914 to 0.665. This result aligns with a recent work demonstrating GraphCodeBERT's vulnerability in cross-language CCD~\cite{mehrotra2023improving}.}
\textcolor{black}{Though cross-language CCD is more challenging, ChatGPT performs better than baselines. We hypothesize that this result could stem from the pre-training and datasets and tasks used to train the LLMs. Note that LLMs have great performance for code translation. In Prompt-2, we ask the model to identify the problem first. So it might use some capabilities for code translation in Java-Ruby pair CCD. However, more investigations are required to determine the reason. }

To further examine ChatGPT's vulnerability to problem complexity, we calculated two metrics assessing the difficulty and complexity of each problem, Acceptance Rate, and Cyclomatic Complexity. 
The former indicates the number of accepted submissions compared to all submissions of a specific problem and the latter is a metric for computing the complexity of a program.
Table \ref{tab:complexity} presents the average of these metrics for the misclassified problems shared between Java-Java and Java-Ruby pairs. In Table \ref{tab:complexity}, positive samples represent the average difficulty metrics for clone pairs related to problems that ChatGPT erroneously labeled as non-clone, while negative samples show the average difficulty for misclassified non-clone pairs. The table's final row calculates the average for all selected CodeNET problems used to sample the datasets in this study i.e., the test set minus the misclassified problems. 
Table \ref{tab:complexity} indicates that the complexity of positive problems is $2.54$, which is less than the average complexity of selected problems in CodeNET. On the other hand, the average complexity for negative samples is $3.16$, which is significantly more than the average complexity. Figure \ref{fig:complexity} provides a clearer illustration of these metrics by comparing the average Acceptance Rate and average Cyclomatic Complexity for positive mislabeled samples and negative misclassified samples, alongside the average values for these metrics in the test set. \textcolor{black}{This result indicates that based on metrics of Acceptance Rate and Cyclomatic Complexity, the complexity of problems influences ChatGPT's performance on CCD. Our hypothesis is that the more complex the code pairs are, the more challenging this reasoning becomes for ChatGPT on code clone detection.}

\textcolor{black}{\textbf{Threats.} We utilized a public high quality dataset and used an schema that is consistent with the benchmarks for dataset creation, and used standard evaluation metrics, to mitigate the internal and construct threats.}
\textcolor{black}{While the approach and methodology may not differ significantly for other programming languages and models, the results might not be generalizable to other languages. }

\section{Conclusions and Future Work}
We presented an initial investigation of applying LLMs for Type-4 code clone detection. The preliminary results are promising as we found that ChatGPT is effective, and its performance is correlated with the problem difficulty and prompt design.
\textcolor{black}{ In the future, we intend to use other LLMs pre-trained on code, like Phi-2 and Mistral, and investigate the reasons of our findings, as well as whether the performances are related to specific programming languages. }

\begin{acks}

This research is supported by a grant from the Natural Sciences and Engineering Research Council of Canada RGPIN-2019-05175.

\end{acks}
\balance

\bibliographystyle{IEEEbib}
\bibliography{refs}

\clearpage

\end{document}